\documentclass[draft]{agujournal}
\draftfalse
\journalname{JGR-Space Physics}
\usepackage{amsmath}
\usepackage{amsfonts}
\usepackage{amssymb}
\usepackage{graphicx}
\usepackage{verbatim}

\begin{document}

\title{Formation of plasma around a small meteoroid: 2. Implications for radar head echo}

\authors{Y.~S.~Dimant\affil{1}\  and M.~M.~Oppenheim\affil{1}}
\affiliation{1}{Center for Space Physics, Boston University}

\correspondingauthor{Y.~S. Dimant}{dimant@bu.edu}

\begin{keypoints}
\item Calculates the spatial distribution of the plasma density around a small ablating meteoroid
\item Plasma density scales with the collisional mean free path and is independent of the meteoroid velocity
\item Provides a basis for realistic modeling of radar head echoes
\end{keypoints}

\begin{abstract}
This paper calculates the spatial distribution of the plasma responsible for
radar head echoes by applying the kinetic theory developed in the companion
paper (Dimant and Oppenheim, arXiv:1608.08524). This
results in a set of analytic expressions for the plasma density as a function
of distance from the meteoroid. It shows that, at distances less than a
collisional mean-free-path from the meteoroid surface, the plasma density
drops in proportion to $1/R$ where $R$ is the distance from the meteoroid
center; and, at distances much longer than the mean-free-path behind the
meteoroid, the density diminishes at a rate proportional to $1/R^{2}$. The
results of this paper should be used for modeling and analysis of radar head echoes.

\end{abstract}

\section{Introduction\label{Introduction}}

The radar head echo is a signal that reflects from the plasma surrounding the
fast-descending meteoroid and is doppler-shifted by approximately the
meteoroid velocity. Only a small volume of the dense plasma sufficiently close
to the meteoroid contributes to the corresponding radar wave reflection.
Quantitative knowledge of the spatial structure of the near-meteoroid plasma
is crucial for accurate modeling the head echo radar reflections
\citep{Bronshten:Physics83,Ceplecha:Meteor1998,Close:NewMethod2005,Campbell-Brown:Meteoroid2007}.

In the companion paper \citep{Dimant:Formation1_17}, hereinafter referred to as
Paper~1, we developed a first-principle kinetic theory of the plasma formed
around a small meteoroid as it moves through the atmosphere at hypersonic
speeds. Using a number of easily justified assumptions, we obtained
approximate analytic expressions describing velocity distributions of meteoric
ions and neutrals. In this paper, we calculate the spatial structure of the
plasma density that follows from the kinetic theory developed in Paper~1. This
calculation demonstrates that this spatial structure differs dramatically from
a simple Gaussian or exponential distribution currently employed for modeling
radar wave scattering from the meteor plasma \citep{Close:NewMethod2005,Marshall:FDTD15}. 
This research does not describe the distribution of
plasma or neutrals in the meteoroid tail where particles lag well behind the
meteoroid after having collided more than once.

Simple analysis of individual collisions between particles indicates that
heavy meteoric particles in the near-meteor sheath consist predominantly of
the `primary' and `secondary' particles. By a primary particle we mean an
ablated meteoroid particle that moves freely with a ballistic trajectory until
it collides with an atmospheric molecule. These primary particles are
predominantly neutral. A secondary particle is a former primary particle that
experienced exactly one collision, either scattering or ionizing. Most of the
near-meteoroid ions responsible for head echoes belong to the group of
secondary particles. The vast majority of ions that experienced multiple
collisions since the original ablation lag behind the fast-moving meteoroid
and form a long-lived extended column of plasma visible to radars through
specular or non-specular echoes.

Given the velocity distributions developed in Paper 1 as a function of spatial
coordinates, one can integrate over velocity variables to find the
corresponding particle density. However, the complexity of these analytic
expressions makes this non-trivial. This paper makes an additional simplifying
assumption about the collision model (the isotropic differential cross-section
of ionization) and then integrates over the velocities to obtain the meteor
gas and plasma density as a function of distance from the meteoroid.

The paper is organized as follows.
Section~\ref{Summary of the ion distribution function} summarizes the results
of Paper~1 on the ion distribution function.
Section~\ref{Plasma density calculations} performs the calculations of the
near-meteoroid plasma density. Section~\ref{Discussion} discusses implications
of our theory and some caveats. Section~\ref{Conclusions} lists the major
underlying assumptions and discusses the paper results.

\section{Summary of the ion distribution
function\label{Summary of the ion distribution function}}

\begin{figure}[h]
\centering
\includegraphics[width=30pc]{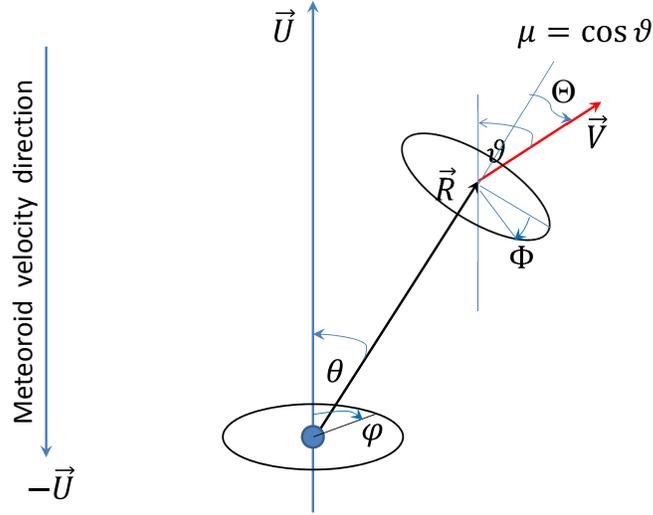}
\caption{Nomenclature of spatial coordinates and velocity variables. The
spatial variables $R=|\vec{R}|$, $\theta$, $\varphi$ denote the radius and two
angles of the spherical coordinate system with the origin at the meteoroid
center and the major axis anti-parallel to the meteoroid velocity (shown on
the left). All other variables pertain to the particle velocity space:
$V=|\vec{V}|$ is the particle speed, $\vartheta$ is the polar angle of
$\vec{V}$ with respect to the local axis parallel to $\vec{U}$, $\Phi$ is the
axial angle measured from the common $\vec{U}$-$\vec{R}$ plane; $\Theta$ is
the polar angle of $\vec{V}$ with respect to the local radial distance
$\vec{R}$.}
\label{Fig:Cartoon_reproduced}
\end{figure}

Paper~1 does all our calculations in the rest frame of a meteoroid moving
through the atmosphere with the local velocity $-\vec{U}$, so that in this
frame the impinging atmospheric particles move with the opposite velocity,
$\vec{U}$. We define the coordinate system with the major axis passing through
the meteoroid center and parallel to $\vec{U}$. Due to the axial symmetry
about $\vec{U}$, we characterize the real space by two spherical coordinates:
the radial distance from the meteoroid center, $R$, and the polar angle,
$\theta$, measured from the major axis ($\theta=0$ corresponds to the major
semi-axis behind the meteoroid, while $\theta=\pi$ corresponds to the opposite
semi-axis in front of it). Figure~\ref{Fig:Cartoon_reproduced}, reproduced
from Paper~1, explains all relevant notations.

The velocity distribution of secondary ions, $f^{(2)}$, is expressed as a
function of three velocity variables that are invariants of the ion
collisionless motion. These variables include the ion speed, $V$, the cosine
of the angle between the ion velocity vector $\vec{V}$ and $\vec{U}$,
$\mu=\cos\vartheta$, and a normalized angular momentum variable, $R_{0}$,
which to the minimum distance between the ion trajectory and the meteoroid
center, $R_{0}=R\sin\Theta$, where $\Theta$ is the polar angle of $\vec{V}$
with respect to the local radius vector $\vec{R}$. The entire set of
velocity-space variables also includes a discrete variable $\sigma_{R}$ which
takes two values, $\pm1$, depending on the sign of the particle radial
velocity,%
\begin{equation}
V_{R}=\frac{dR}{dt}\equiv V\cos\Theta=\sigma_{R}\sqrt{1-\frac{R_{0}^{2}}%
{R^{2}}}\ V. \label{V_r_snova}%
\end{equation}
The value of $\sigma_{R}=+1$ corresponds to the outgoing particles, $V_{R}>0$,
while $\sigma_{R}=-1$ corresponds to the incoming particles, $V_{R}<0$. At any
location, the entire distribution function is given by a sum of the two
corresponding functions,%
\begin{equation}
f^{(2)}(V,\mu,R_{0};R,\theta)=\sum_{\sigma_{R}=\pm1}f_{\sigma_{R}}^{(2)}%
(V,\mu,R_{0};R,\theta). \label{sum_f}%
\end{equation}
The functions $f_{\sigma_{R}}^{(2)}$ are non-zero provided $\mu=\cos
\vartheta>0$; otherwise $f_{\sigma_{R}}^{(2)}=0$,%
\begin{align}
&  \left.  f_{\sigma_{R}}^{(2)}(V,R,\theta)\right\vert _{\mu>0}=L_{\sigma_{R}%
}\ \delta\left(  V-\frac{2m_{\beta}\mu U}{m+m_{\beta}}\right)  ,\nonumber\\
L_{\sigma_{R}}  &  =\frac{G_{\mathrm{ion}}(U,1-2\mu^{2})n_{0}n_{\mathrm{A}}%
}{\sqrt{3}\ \mu U^{2}}\left(  1+\frac{m}{m_{\beta}}\right)  ^{3}%
I(R,R_{0}),\label{f^(2)_via_I}\\
&  \left.  f_{\sigma_{R}}^{\left(  2\right)  }(V,R,\theta)\right\vert _{\mu
<0}=0.\nonumber
\end{align}
The quantities $n_{0}$ and $n_{\mathrm{A}}$ are the densities of the ablated
particles at the meteoroid surface and of the atmospheric particles at a given
altitude, respectively. The quantity $G_{\mathrm{ion}}(U,1-2\mu^{2})$
originates from the differential cross-section of ionizing collisions,
$G_{\mathrm{ion}}$, expressed as a function of the relative speed between the
two colliding particles, $u=\left\vert \vec{u}\right\vert $, and the cosine of
the scattering angle, $\Theta_{\mathrm{sc}}$. In this paper, we simplify our
treatment by assuming $G_{\mathrm{ion}}$ to be a function of only $u\approx
U$. The corresponding angular dependence in the relevant energy range is
generally unknown, but the assumption of isotropic $G_{\mathrm{ion}}(U)$ is reasonable.

The condition $\mu>0$ is fulfilled if either
\begin{subequations}
\label{sgn}%
\begin{align}
&  \sigma_{R}=\mathrm{sgn}(\cos\theta)\qquad\text{and}\qquad0<R_{0}%
<R_{c}(\theta,\Phi)\label{sigma=sgn}\\
\text{or}  & \nonumber\\
&  \sigma_{R}=-\mathrm{sgn}(\cos\theta)\qquad\text{and}\qquad R_{c}%
(\theta,\Phi)<R_{0}<R, \label{sigma=-sgn}%
\end{align}
where $\mathrm{sgn}(x)$ means the sign of $x$ and%
\end{subequations}
\begin{equation}
R_{c}(\theta,\Phi)=\frac{R\left\vert \cos\theta\right\vert }{\sqrt{1-\sin
^{2}\theta\sin^{2}\Phi}}. \label{R_c(Phi)}%
\end{equation}
Here $\Phi$ is the axial angle of the particle velocity $\vec{V}$ around the
direction of the local radius-vector $\vec{R}$ (see
Figure~\ref{Fig:Cartoon_reproduced}) and we set the origin $\Phi=0$ where
$\vec{V}$ lies in the common $\vec{R}$-$\vec{U}$ plane.

In this paper, we consider the meteor plasma located not too close to the
meteoroid, $R\gg r_{\mathrm{M}}$. The corresponding multiplier $I(R,R_{0})$ in
equation~(\ref{f^(2)_via_I}) has the following piecewise definition:%
\begin{equation}
I(R,R_{0})=\left\{
\begin{array}
[c]{ccc}%
J_{R}^{\infty} & \text{for} & \sigma_{R}=-1,\\
&  & \\
J_{R_{0}}^{\infty}+J_{R_{0}}^{R}=2J_{R_{0}}^{\infty}-J_{R}^{\infty} &
\text{for} & \sigma_{R}=+1,
\end{array}
\right.  \label{III}%
\end{equation}
where, under constraints of $R,R_{0}>3r_{\mathrm{M}}$, the well-convergent
integral $J_{a}^{b}$, taken as a function of its integration limits, $b>a\geq
R_{0}$, is given by%
\begin{equation}
J_{a}^{b}\approx r_{\mathrm{M}}^{2}\int_{a}^{b}\left[  1+\left(
\frac{R^{\prime}}{\lambda_{T}^{(1)}}\right)  ^{2/3}\right]  \exp\left[
-\ \frac{3}{2}\left(  \frac{R^{\prime}}{\lambda_{T}^{(1)}}\right)
^{2/3}\right]  \frac{dR^{\prime}}{R^{\prime}\sqrt{\left(  R^{\prime}\right)
^{2}-R_{0}^{2}}}. \label{J_a^b_R>3epsilon}%
\end{equation}
Here $\lambda_{T}^{(1)}$ is the mean free path of the primary (ablated)
particles,%
\begin{equation}
\lambda_{T}^{(1)}=\frac{V_{T}}{\nu_{T}^{(1)}},\qquad V_{T}=\left(  \frac
{T_{M}}{m_{\mathrm{M}}}\right)  ^{1/2},\qquad\nu_{T}^{(1)}\approx2\pi
n_{\mathrm{A}}U\int_{-1}^{1}G^{\left(  1\right)  }(U,\Lambda)d\Lambda,
\label{lambda_T^(1)}%
\end{equation}
where $T_{M}$ and $m_{\mathrm{M}}$ are the temperature and mass of the primary
meteor particles. The quantity $G^{\left(  1\right)  }(U,\Lambda)$ includes
all collisions that result in scattering of the primary neutral particles. The
expression for $G^{\left(  1\right)  }$, as that for $G_{\mathrm{ion}}$, takes
into account that $V_{T}\ll U$, so that the collision frequency $\nu_{T}%
^{(1)}$ depends only on the meteoroid speed, $U$, and hence is the same for
all particles. This reduces $\lambda_{T}^{(1)}$ to a constant value which
becomes the characteristic length-scale of the near-meteoroid plasma.

The general integral $J_{a}^{b}$ cannot be calculated exactly, but the
particular integral $J_{R_{0}}^{\infty}$ has an almost perfect analytic
approximation,%
\begin{equation}
J_{R_{0}}^{\infty}\approx\frac{\pi r_{\mathrm{M}}^{2}}{2R_{0}}\sqrt{1+\frac
{2}{\pi}\left(  \frac{R_{0}}{\lambda_{T}^{(1)}}\right)  ^{2/3}}\ \exp\left[
-\ \frac{3}{2}\left(  \frac{R_{0}}{\lambda_{T}^{(1)}}\right)  ^{2/3}\right]  ,
\label{J_R_0_interpo}%
\end{equation}
accurate within $1\%$ for all $R_{0}$. As we demonstrate below, depending on
the specific calculation, it may become beneficial to use either the exact
original integral expression for $J_{a}^{b}$ given by
equation~(\ref{J_a^b_R>3epsilon}) or (only for $J_{R_{0}}^{\infty}$) its
approximation given by equation~(\ref{J_R_0_interpo}).

For local calculations of the ion density it is more convenient to pass from
the invariant velocity variables $V,R_{0},\mu$ to local variables
$V,R_{0},\Phi$, where $\Theta$ and $\Phi$ are the polar and axial angles of
the ion velocity about the direction of the local radius-vector $\vec{R}$, as
depicted by Figure~\ref{Fig:Cartoon_reproduced}.

\section{Plasma density calculations\label{Plasma density calculations}}

Radar head echo is determined by the spatial distribution of the electron
density around the meteoroid. The near-meteoroid plasma is quasi-neutral, so
that the electron density almost equals that of ions, $n_{e}\approx n_{i}=n$.
We calculate the spatial distribution of the ion density based on the
distribution function explained in
section~\ref{Summary of the ion distribution function}.

The ion density can be easily calculated in the far region of $R\gg\lambda
_{T}^{(1)}$. Albeit less simple, but $n^{(2)}(R,\theta)$ can also be
explicitly calculated in the opposite limit of $R\ll\lambda_{T}^{(1)}$. In the
entire space of arbitrary $R$, we were unable to find a unified purely
algebraic expression for $n^{(2)}(R,\theta)$. However, we have reduced the
general 3D velocity-space integral to a much simpler expression for $n^{(2)}$
in terms of normalized variables and parameters, as explained below. This
universal expression involves only treatable analytical functions and two 1D
integral functions suitable for simple numerical integration and tabulation.
The resultant universal expression for $n^{(2)}$ makes the future analysis and
computer modeling of the radar signal much easier.

\subsection{Preliminary remarks}

At a given location determined by $R$ and $\theta$, the ion density is given
by $n^{\left(  2\right)  }=\sum_{\sigma_{R}=\pm1}\int f_{\sigma_{R}}%
^{(2)}V^{2}dVd\Omega$, where $d\Omega=d\left(  \cos\Theta\right)  d\Phi$
denotes the elementary volume of the local solid angle. Choosing instead of
$\Theta$ the new variable $R_{0}=R\sin\Theta$, we obtain%
\begin{equation}
n^{\left(  2\right)  }=\frac{1}{R}\sum_{\sigma_{R}=\pm1}\int_{0}^{R}%
\frac{R_{0}dR_{0}}{\sqrt{R^{2}-R_{0}^{2}}}\int_{0}^{2\pi}d\Phi\int_{0}%
^{\infty}f_{\sigma_{R}}^{(2)}V^{2}dV. \label{n^(2)_snova}%
\end{equation}

First, we integrate over $V$ to eliminate the $\delta$-function in equation
(\ref{f^(2)_via_I}),%
\begin{equation}
\int_{0}^{\infty}f_{\sigma_{R}}^{(2)}V^{2}dV=\frac{4\mu G_{\mathrm{ion}}%
n_{0}n_{\mathrm{A}}}{\sqrt{3}}\left(  1+\frac{m}{m_{\beta}}\right)
I(R,R_{0}). \label{yields}%
\end{equation}
As a result of this simple integration, the previously singular factor $\mu$
in the expression for $L_{\sigma_{R}}$ has moved from its denominator to the
numerator, reducing dramatically the relative contribution of the
`small-angle' ($\Theta_{\mathrm{sc}}=1-2\mu^{2}\approx1$, $\mu\ll1$)
ionization where some assumptions of our general theory are invalid
\citep{Dimant:Formation1_17}. Since we assumed above isotropic $G_{\mathrm{ion}%
}=G_{\mathrm{ion}}(U)$, the only $\Phi$-dependent quantity in the right-hand
side (RHS) of equation (\ref{yields}) is $\mu$ in the numerator. This variable
is expressed in terms of $R_{0}$ and $\Phi$ as%
\begin{equation}
\mu=\sigma_{R}\sqrt{1-\frac{R_{0}^{2}}{R^{2}}}\cos\theta+\frac{R_{0}\sin
\theta}{R}\ \cos\Phi. \label{mumumu}%
\end{equation}
The function $I(R,R_{0})$, along with the corresponding integral expressions
for $J_{a}^{b}$, are described by equations~(\ref{III}) to
(\ref{J_R_0_interpo}).

\subsection{Long-distance asymptotics, $R\gg\lambda_{T}^{(1)}$, behind the
meteoroid\label{Long-distance asymptotics}}

We start by calculating the ion density at the simplest limit of long radial
distances, $R\gg\lambda_{T}^{(1)}$, behind the meteoroid. Ignoring
exponentially small densities (as explained below), we will consider only the
space behind the meteoroid, $\cos\theta>0$. Outgoing particles within the
dominant beam-like (along $\vec{R}$) velocity distribution make the major
contribution to $n^{\left(  2\right)  }$. In equations (\ref{n^(2)_snova}) and
(\ref{yields}), setting $R_{0}\ll R$, $\vartheta\approx\theta$, $\mu
\approx\cos\theta$, while neglecting the exponentially small quantity
$J_{R}^{\infty}$ in equation~(\ref{III}), we obtain $I(R,R_{0})\approx
2J_{R_{0}}^{\infty}$. This allows us to easily integrate the RHS of
equation~(\ref{n^(2)_snova}) over $\Phi$. In the exact integral expression
given by (\ref{J_a^b_R>3epsilon}), the primary contribution to $J_{R_{0}%
}^{\infty}$ arises from components of $f_{+}^{(2)}$ near the meteoroid,
$R^{\prime}\lesssim\lambda_{T}^{(1)}\ll R$. This allows us to extend the upper
integration limit to infinity. This yields%
\begin{align}
&  \left.  n^{\left(  2\right)  }\right\vert _{R\gg\lambda_{T}^{(1)}}%
\approx\frac{16\pi r_{\mathrm{M}}^{2}G_{\mathrm{ion}}n_{0}n_{\mathrm{A}}%
\cos\theta}{R^{2}\sqrt{3}}\left(  1+\frac{m}{m_{\beta}}\right) \nonumber\\
&  \times\int_{0}^{\infty}R_{0}dR_{0}\int_{R_{0}}^{\infty}\left[  1+\left(
\frac{R^{\prime}}{\lambda_{T}^{(1)}}\right)  ^{2/3}\right]  \exp\left[
-\ \frac{3}{2}\left(  \frac{R^{\prime}}{\lambda_{T}^{(1)}}\right)
^{2/3}\right]  \frac{dR^{\prime}}{R^{\prime}\sqrt{\left(  R^{\prime}\right)
^{2}-R_{0}^{2}}}. \label{n^(2)_long_start}%
\end{align}
Changing here the order of integration with the corresponding adjustment of
the integration limits, $\int_{0}^{\infty}dR_{0}\int_{R_{0}}^{\infty}\left(
\cdots\right)  dR^{\prime}=\int_{0}^{\infty}dR^{\prime}\int_{0}^{R^{\prime}%
}\left(  \cdots\right)  dR_{0}$, and using the simple identities
\begin{align*}
\int_{0}^{R^{\prime}}\frac{R_{0}dR_{0}}{\sqrt{\left(  R^{\prime}\right)
^{2}-R_{0}^{2}}}  &  =R^{\prime},\\
\int_{0}^{\infty}\left[  1+\left(  \frac{R^{\prime}}{\lambda_{T}^{(1)}%
}\right)  ^{2/3}\right]  \exp\left[  -\ \frac{3}{2}\left(  \frac{R^{\prime}%
}{\lambda_{T}^{(1)}}\right)  ^{2/3}\right]  dR^{\prime}  &  =\sqrt{\frac{2\pi
}{3}}\ \lambda_{T}^{(1)},
\end{align*}
we obtain for $\cos\theta>0$:
\begin{equation}
\left.  n^{\left(  2\right)  }\right\vert _{R\gg\lambda_{T}^{(1)}}\approx
\frac{kr_{\mathrm{M}}^{2}\lambda_{T}^{(1)}G_{\mathrm{ion}}n_{0}n_{\mathrm{A}}%
}{R^{2}}\left(  1+\frac{m}{m_{\beta}}\right)  \cos\theta, \label{n_R>>lambda}%
\end{equation}
where $k=16\sqrt{2}\ \pi^{\frac{3}{2}}/3$. Equation~(\ref{n_R>>lambda}) shows
that at $R\gg\lambda_{T}^{(1)}$ the density of the major ion population fall
off as $(\cos\theta)/R^{2}$.

If, however, instead of using the exact integral expression for $J_{R_{0}%
}^{\infty}$ given by equation (\ref{J_a^b_R>3epsilon}), we employed its
approximation given by equation~(\ref{J_R_0_interpo}) then, applying the
identity
\begin{equation}
\int_{0}^{\infty}\sqrt{\left(  1+\frac{4x}{3\pi}\right)  x}\ \exp\left(
-\ x\right)  dx=\frac{\sqrt{3\pi}}{4}\exp\left(  \frac{3\pi}{8}\right)
K_{1}\left(  \frac{3\pi}{8}\right)  , \label{K_1}%
\end{equation}
deduced from \citep[][equation~3.372]{Gradshteyn:Table94}, with the modified
Bessel function of the second kind $K_{\nu}(x)$, we would obtain
equation~(\ref{n_R>>lambda}) with $k$ replaced by a formally different
coefficient, $k_{1}=(2^{\frac{3}{2}}\pi^{\frac{5}{2}}/\sqrt{3})\exp\left(
3\pi/8\right)  K_{1}\left(  3\pi/8\right)  $. However, the numerical values of
$k\approx42$ and $k_{1}\approx41.74$ are so close to each other that can be
considered as essentially the same. This confirms that the approximate
expression $J_{R_{0}}^{\infty}$ given by equation~(\ref{J_R_0_interpo}) is
reasonably accurate and can be successfully used in other occasions, as done below.

\subsection{General distances \label{Moderate distances}}

For all but largest meteors the radar head echo is formed within moderate
radial distances of $R\sim\lambda_{T}^{(1)}$, the most difficult domain to
treat analytically. Below, we reduce the general expression for $n^{\left(
2\right)  }$ to a simpler form, more suitable for a further analytic or
numerical treatment, and then obtain the explicit spatial distribution of
$n^{\left(  2\right)  }$ for $R\ll\lambda_{T}^{(1)}$, the limit opposite to
that considered in section~\ref{Long-distance asymptotics}. After that, we
will discuss the general case, using numerical integrations.

\subsubsection{Reduction of the general ion density \label{Reduction}}

Under assumption of the isotropic differential cross-section, $G_{\mathrm{ion}%
}(U)$, equation~(\ref{yields}) involves $\mu$ only as a linear multiplier. For
the further analysis, equation~(\ref{n^(2)_snova}) with the integration over
$\Phi$ is no longer convenient. More advantageous is integrating over $\mu$,
where $\mu$ is given by equation~(\ref{mumumu}). Introducing a dimensionless
variable%
\begin{equation}
\xi_{0}=\frac{R_{0}}{R}=\sin\Theta\leq1 \label{xi_0}%
\end{equation}
and changing variables $R_{0},\Phi$ to $\xi_{0},\mu$, we arrive at
\begin{subequations}
\label{n^(2),I_muha}%
\begin{align}
&  n^{\left(  2\right)  }=2\times\frac{4n_{0}n_{\mathrm{A}}}{\sqrt{3}}\left(
1+\frac{m}{m_{\beta}}\right)  G_{\mathrm{ion}}M\label{n^(2)_prome_2}\\
&  M\equiv\sum_{\sigma_{R}=\pm1}\int_{0}^{1}\frac{I(R,\xi_{0}R)\xi_{0}d\xi
_{0}}{\sqrt{1-\xi_{0}^{2}}}\ I_{\mu}(\xi_{0}),\label{M}\\
I_{\mu}(\xi_{0})  &  =\int\frac{\mathrm{H}\left(  \mu\right)  \mu d\mu}%
{\sqrt{\xi_{0}^{2}\sin^{2}\theta-\left(  \mu-\sigma_{R}\sqrt{1-\xi_{0}^{2}%
}\cos\theta\right)  ^{2}}}, \label{I_muha}%
\end{align}
where the function $I(R,R_{0})$ is given by equation~(\ref{III}) and
$\mathrm{H}\left(  x\right)  $ is the Heaviside step-function ($\mathrm{H}%
\left(  x\right)  =1$ for $x\geq0$ and $\mathrm{H}\left(  x\right)  =0$ for
$x<0$). The latter takes into account the fact that the distribution function
of secondary particles is non-zero only for positive $\mu$, as discussed in
Paper~1. The integration in $I_{\mu}(\xi_{0})$ is performed over the entire
$\mu$-range where the expression under the square root is non-negative. The
factor `$2$' in front of the RHS of equation~(\ref{n^(2)_prome_2}) takes into
account the fact that each value of $\mu$ corresponds to two symmetric values
of $\Phi$ with the same $\cos\Phi$ but opposite $\sin\Phi$. Introducing%
\end{subequations}
\begin{equation}
\mu_{1}=\sigma_{R}\sqrt{1-\xi_{0}^{2}}\cos\theta-\xi_{0}\sin\theta,\qquad
\mu_{2}=\sigma_{R}\sqrt{1-\xi_{0}^{2}}\cos\theta+\xi_{0}\sin\theta,
\label{mu_1,2}%
\end{equation}
and eliminating the step-function, we can rewrite $I_{\mu}(\xi_{0})$ in
equation~(\ref{I_muha}) as%
\begin{equation}
I_{\mu}(\xi_{0})=\int_{\max(\mu_{1},0)}^{\max(\mu_{2},0)}\frac{\mu d\mu}%
{\sqrt{\left(  \mu_{2}-\mu\right)  \left(  \mu-\mu_{1}\right)  }},
\label{I_mu}%
\end{equation}
where the integration limits take into account that in general case $\mu
_{1,2}$ can be negative. For all signs of $\cos\theta$, the relations between
$\mu_{1,2}$ and $0$ are listed in this table:%
\begin{equation}%
\begin{tabular}
[c]{|l|l|l|}\hline
& $\xi_{0}<\left\vert \cos\theta\right\vert $ & $\xi_{0}>\left\vert \cos
\theta\right\vert $\\\hline
$\sigma_{R}\cos\theta<0$ & $\mu_{1}<\mu_{2}<0$ & $\mu_{1}<0<\mu_{2}$\\\hline
$\sigma_{R}\cos\theta>0$ & $\mu_{2}>\mu_{1}>0$ & $\mu_{1}<0<\mu_{2}$\\\hline
\end{tabular}
\ \ \ \ \ . \label{Table:mu>>}%
\end{equation}
All this yields for $I_{\mu}$, defined by equation~(\ref{I_muha}), a
piece-wise expression:%
\begin{equation}
I_{\mu}(\xi_{0})=\left\{
\begin{array}
[c]{ccc}%
\int_{\mu_{1}}^{\mu_{2}}\frac{\mu d\mu}{\sqrt{\left(  \mu_{2}-\mu\right)
\left(  \mu-\mu_{1}\right)  }} & \text{if} & \xi_{0}<\left\vert \cos
\theta\right\vert \text{ and }\sigma_{R}\cos\theta>0,\\
\int_{0}^{\mu_{2}}\frac{\mu d\mu}{\sqrt{\left(  \mu_{2}-\mu\right)  \left(
\mu-\mu_{1}\right)  }} & \text{if} & \xi_{0}>\left\vert \cos\theta\right\vert
\text{ for all }\sigma_{R}\cos\theta,\\
0 & \text{if} & \xi_{0}<\left\vert \cos\theta\right\vert \text{ and }%
\sigma_{R}\cos\theta<0.
\end{array}
\right.  . \label{I_mu_proma}%
\end{equation}
Using the definitions given by equation~(\ref{mu_1,2}), we obtain:%
\begin{equation}
\int_{\mu_{1}}^{\mu_{2}}\frac{\mu d\mu}{\sqrt{\left(  \mu_{2}-\mu\right)
\left(  \mu-\mu_{1}\right)  }}=\pi\sigma_{R}\sqrt{1-\xi_{0}^{2}}\cos\theta,
\label{Int_mu_full_again}%
\end{equation}
and (for $\xi_{0}>\left\vert \cos\theta\right\vert $):%
\begin{align}
&  \int_{0}^{\mu_{2}}\frac{\mu d\mu}{\sqrt{\left(  \mu_{2}-\mu\right)  \left(
\mu-\mu_{1}\right)  }}\nonumber\\
&  =\sigma_{R}\left(  \frac{\pi}{2}+\arcsin\frac{\sigma_{R}\sqrt{1-\xi_{0}%
^{2}}\cos\theta}{\xi_{0}\sin\theta}\right)  \sqrt{1-\xi_{0}^{2}}\cos
\theta+\sqrt{\xi_{0}^{2}-\cos^{2}\theta}. \label{Int_mu_partial_again}%
\end{align}
Recalling equation~(\ref{III}), for the quantity $M$, defined by
equation~(\ref{M}), we obtain%
\begin{align}
M  &  =\int_{0}^{1}\frac{\left(  2J_{R_{0}}^{\infty}-J_{R}^{\infty}\right)
_{\sigma_{R>0}}\xi_{0}d\xi_{0}}{\sqrt{1-\xi_{0}^{2}}}\int_{\max(\mu_{1}%
,0)}^{\max(\mu_{2},0)}\frac{\mu d\mu}{\sqrt{\left(  \mu_{2}-\mu\right)
\left(  \mu-\mu_{1}\right)  _{\sigma_{R=+1}}}}\label{K}\\
&  +\int_{0}^{1}\frac{\left(  J_{R}^{\infty}\right)  _{\sigma_{R<0}}\xi
_{0}d\xi_{0}}{\sqrt{1-\xi_{0}^{2}}}\int_{\max(\mu_{1},0)}^{\max(\mu_{2}%
,0)}\frac{\mu d\mu}{\sqrt{\left(  \mu_{2}-\mu\right)  \left(  \mu-\mu
_{1}\right)  _{\sigma_{R=-1}}}}\nonumber
\end{align}
Regrouping the terms, and using equations~(\ref{I_mu_proma}%
)--(\ref{Int_mu_partial_again}), for all $\theta$ we obtain%
\begin{align}
&  M=\pi\left\vert \cos\theta\right\vert \int_{0}^{\left\vert \cos
\theta\right\vert }J_{R_{0}}^{\infty}\xi_{0}d\xi_{0}+\left(  \pi\cos
\theta\right)  \int_{0}^{1}J_{R_{0}}^{R}\xi_{0}d\xi_{0}\nonumber\\
&  +2\int_{\left\vert \cos\theta\right\vert }^{1}J_{R_{0}}^{\infty}\xi
_{0}\sqrt{\frac{\xi_{0}^{2}-\cos^{2}\theta}{1-\xi_{0}^{2}}}\ d\xi
_{0}\label{K_snova}\\
&  +2\left\vert \cos\theta\right\vert \int_{\left\vert \cos\theta\right\vert
}^{1}J_{R_{0}}^{\infty}\xi_{0}\arcsin\frac{\sqrt{1-\xi_{0}^{2}}\left\vert
\cos\theta\right\vert }{\xi_{0}\sin\theta}\ d\xi_{0}\ ,\nonumber
\end{align}
where $J_{R_{0}}^{R}=J_{R_{0}}^{\infty}-J_{R}^{\infty}$. All terms in the RHS
of equation~(\ref{K_snova}) are symmetric with respect to the sign of
$(\cos\theta)$, except the second term which is antisymmetric. This term is
responsible for the entire asymmetry between the locations in front of the
meteoroid ($\cos\theta<0$) and behind it ($\cos\theta>0$).

To simplify further, we introduce other variables and parameters,%
\begin{equation}
\eta=\frac{R^{\prime}}{R},\qquad\beta=\left(  \frac{R}{\lambda_{T}^{(1)}%
}\right)  ^{\frac{2}{3}},\qquad q=\frac{r_{\mathrm{M}}^{2}}{R},
\label{beta_snova}%
\end{equation}
where $R^{\prime}$ is the integration variable in $J_{a}^{b}$, defined by
equation~(\ref{J_a^b_R>3epsilon}). With use of these dimensionless quantities,
equation~(\ref{J_a^b_R>3epsilon}) yields
\begin{subequations}
\label{J_R_0^infty,1}%
\begin{align}
J_{R_{0}}^{\infty}  &  =q\int_{\xi_{0}}^{\infty}\left(  1+\beta\eta^{\frac
{2}{3}}\right)  \exp\left(  -\ \frac{3\beta\eta^{\frac{2}{3}}}{2}\right)
\frac{d\eta}{\eta\sqrt{\eta^{2}-\xi_{0}^{2}}},\label{J_R_0^infty}\\
J_{R_{0}}^{R}  &  =q\int_{\xi_{0}}^{1}\left(  1+\beta\eta^{\frac{2}{3}%
}\right)  \exp\left(  -\ \frac{3\beta\eta^{\frac{2}{3}}}{2}\right)
\frac{d\eta}{\eta\sqrt{\eta^{2}-\xi_{0}^{2}}}. \label{J_R_0^1}%
\end{align}
For some calculations, we will also need approximate
equation~(\ref{J_R_0_interpo}) for $J_{R_{0}}^{\infty}$,%
\end{subequations}
\begin{equation}
J_{R_{0}}^{\infty}\approx\frac{\pi q}{2\xi_{0}}\sqrt{1+\frac{2}{\pi}\ \beta
\xi_{0}^{\frac{2}{3}}}\exp\left(  -\ \frac{3}{2}\ \beta\xi_{0}^{\frac{2}{3}%
}\right)  . \label{J_R_)_approx_dimensionless}%
\end{equation}
For $J_{R_{0}}^{R}$ we need no approximations, as will become clear soon.

Before proceeding, we check that the long-distance limit of $R\gg\lambda
_{T}^{(1)}$ (i.e., $\beta\gg1$) for the above equations provides a smooth
transition to the range of long distances considered in
section~\ref{Long-distance asymptotics}. Beyond a narrow vicinity around
$\theta=\pi/2$, given by $\left\vert \cos\theta\right\vert \lesssim
\beta^{-3/2}$, we can easily see that in the RHS of equation~(\ref{K_snova})
the third and fourth terms are exponentially small. Neglecting them and
extending the same accuracy to the upper integration limit in the first and
second terms, $J_{R_{0}}^{R}\approx J_{R_{0}}^{\infty}$, we obtain
\begin{equation}
\left.  M\right\vert _{\beta\gg1}\approx\left\{
\begin{array}
[c]{ccc}%
2\pi\left(  \cos\theta\right)  \int_{0}^{\infty}J_{R_{0}}^{\infty}\xi_{0}%
d\xi_{0} & \text{if} & \cos\theta>0,\\
&  & \\
0 & \text{if} & \cos\theta<0.
\end{array}
\right.  \label{K_beta>>1}%
\end{equation}
Using for $J_{R_{0}}^{\infty}$ the exact equation~(\ref{J_R_0^infty}) and
applying for the double integration the same approach as in
section~\ref{Long-distance asymptotics}, we obtain
\[
\left.  M\right\vert _{\beta\gg1}\approx\frac{2\pi q\cos\theta}{\beta
^{\frac{3}{2}}}\sqrt{\frac{2\pi}{3}}.
\]
Returning from the temporary dimensionless parameters $\beta,q$ to the
original coordinate $R$ and inserting the corresponding $M$ to
equation~(\ref{n^(2)_prome_2}), for $\cos\theta>0$ we recreate
equation~(\ref{n_R>>lambda}).

\subsubsection{Short distances, $R\ll\lambda_{T}^{(1)}$%
\label{Very short distances}}

Now we consider the short-distance limit of $R\ll\lambda_{T}^{(1)}$ ($\beta
\ll1$) which is opposite to that discussed in
section~\ref{Long-distance asymptotics}. For not too large integration
variables $R^{\prime}$, $\eta=R^{\prime}/R\ll\beta^{-3/2}$, all factors with
$\beta\eta^{\frac{2}{3}}$ in equation~(\ref{J_R_0^infty,1}) can be neglected.
Since this range of $R^{\prime}$ makes the dominant contribution to all
integrals, we extend this approximation to the entire range of $\eta$, so
that
\begin{subequations}
\label{J_R,R_0_reduced}%
\begin{align}
J_{R_{0}}^{\infty} &  \approx q\int_{\xi_{0}}^{\infty}\frac{d\eta}{\eta
\sqrt{\eta^{2}-\xi_{0}^{2}}}=\frac{\pi q}{2\xi_{0}},\label{J_R_0_reduced}\\
J_{R_{0}}^{R} &  \approx q\int_{\xi_{0}}^{1}\frac{d\eta}{\eta\sqrt{\eta
^{2}-\xi_{0}^{2}}}=\frac{q}{\xi_{0}}\arccos\xi_{0}.\label{J_R_reduced}%
\end{align}
In this limit, the first two terms in the RHS\ of equation~(\ref{K_snova}) can
be easily integrated, yielding $\pi q[\pi(\cos^{2}\theta)/2+\cos\theta]$. The
two remaining terms can be expressed in terms of the complete elliptic
integrals of the 1st and 2nd kind,%
\end{subequations}
\begin{equation}
\mathrm{K}\left(  k\right)  =\int_{0}^{1}\frac{dt}{\sqrt{\left(
1-t^{2}\right)  \left(  1-k^{2}t^{2}\right)  }},\qquad\mathrm{E}\left(
k\right)  =\int_{0}^{1}\sqrt{\frac{1-k^{2}t^{2}}{1-t^{2}}}%
\ dt,\label{F,E_complete}%
\end{equation}
respectively, where $0\leq k<1$. Indeed, for constant $\xi_{0}J_{R_{0}%
}^{\infty}$, as in equation~(\ref{J_R_0_reduced}), the third term in
equation~(\ref{K_snova}) becomes proportional to%
\[
I_{1}=\int_{\left\vert \cos\theta\right\vert }^{1}\sqrt{\frac{\xi_{0}^{2}%
-\cos^{2}\theta}{1-\xi_{0}^{2}}}\ d\xi_{0}.
\]
This integral already resembles an elliptic integral, but reducing $I_{1}$ to
those with the real arguments requires additional efforts. Substituting
$\xi_{0}=\sqrt{1-z^{2}\sin^{2}\theta}$, we reduce $I_{1}$ to $\mathrm{E}%
\left(  \sin\theta\right)  -\left(  \cos^{2}\theta\right)  \mathrm{K}\left(
\sin\theta\right)  $. In a similar way, we can also calculate the fourth term
in equation~(\ref{K_snova}). With constant $J_{R_{0}}^{\infty}\xi_{0}$, this
term becomes proportional to%
\[
I_{2}=\int_{\left\vert \cos\theta\right\vert }^{1}\arcsin\frac{\sqrt{1-\xi
_{0}^{2}}\left\vert \cos\theta\right\vert }{\xi_{0}\sin\theta}\ d\xi_{0}.
\]
Integration of the corresponding indefinite integral by parts gives%
\begin{align*}
&  \int\arcsin\frac{\sqrt{1-\xi_{0}^{2}}\left\vert \cos\theta\right\vert }%
{\xi_{0}\sin\theta}\ d\xi_{0}\\
&  =\xi_{0}\arcsin\frac{\sqrt{1-\xi_{0}^{2}}\left\vert \cos\theta\right\vert
}{\xi_{0}\sin\theta}+\left\vert \cos\theta\right\vert \int\frac{d\xi_{0}%
}{\sqrt{\left(  1-\xi_{0}^{2}\right)  \left(  \xi_{0}^{2}-\cos^{2}%
\theta\right)  }}.
\end{align*}
Making\ the same substitution for the remaining integral in the RHS as done
for $I_{1}$ and evaluating everything over the proper integration limits, we
obtain $I_{2}=(\mathrm{K}\left(  \sin\theta\right)  -\pi/2)|\cos\theta|$. When
adding all terms in the RHS of equation~(\ref{K_snova}), the $\mathrm{K}%
\left(  \sin\theta\right)  $-terms in $I_{1,2}$ cancel and equation~(\ref{M})
reduces to a simple expression: $M\approx\pi q[\cos\theta+\mathrm{E}\left(
\sin\theta\right)  ]$. As a result, the ion density in the short-distance
limit reduces to%
\begin{equation}
\left.  n^{\left(  2\right)  }\right\vert _{R\ll\lambda_{T}^{(1)}}\approx
\frac{8\pi r_{\mathrm{M}}^{2}G_{\mathrm{ion}}n_{0}n_{\mathrm{A}}}{\sqrt{3}%
\,R}\left(  1+\frac{m}{m_{\beta}}\right)  \left[  \cos\theta+\mathrm{E}\left(
\sin\theta\right)  \right]  .\label{n_R<<lambda}%
\end{equation}

Comparing equation~(\ref{n_R<<lambda}) with the opposite limit given by
equation~(\ref{n_R>>lambda}) shows that the $1/R^{2}$-dependency of
$n^{\left(  2\right)  }|_{R\gg\lambda_{T}^{(1)}}$ transforms to the
$1/R$-dependency for $n^{\left(  2\right)  }|_{R\ll\lambda_{T}^{(1)}}$. The
angular $\theta$-dependency also changes significantly. While in the
long-distance limit of $R\gg\lambda_{T}^{(1)}$ ions occupy almost entirely the
half-space behind the meteoroid ($0\leq\theta<\pi/2$), in the short-distance
limit of $R\ll\lambda_{T}^{(1)}$ ions show a noticeable presence in front of
the meteoroid ($\pi/2\leq\theta\leq\pi$) as well. Red dashed curves in
Figure~\ref{Fig:M_versus_theta} show the corresponding angular dependencies
normalized to their maximum values at $\theta=0$.

\subsubsection{Arbitrary distances \label{General case}}

The case of moderate distances $R\sim\lambda_{T}^{(1)}$ is covered by general
equations (\ref{n^(2),I_muha}) and (\ref{K_snova}) with $J_{R_{0}}^{\infty}$
and $J_{R_{0}}^{R}$ expressed in the original integral form by
equation~(\ref{J_R_0^infty,1}), or in an approximate, but explicit, form for
$J_{R_{0}}^{\infty}$ by equation~(\ref{J_R_)_approx_dimensionless}). Unlike
$J_{R_{0}}^{\infty}$, the integral $J_{R_{0}}^{R}$ is involved only in the
second term in the RHS of equation~(\ref{K_snova}). As we show below, this
term can be calculated exactly by using the integral form of
equation~(\ref{J_R_0^1}). Below we obtain the explicit analytic expressions
for the first and second terms in the RHS of equation~(\ref{K_snova}). Being
unable to obtain a general analytic approximation for the two last integral
terms, we will integrate them numerically.

\paragraph{First term in\ equation~(\ref{K_snova}).}

Using~equation~(\ref{J_R_0^infty}), for the integral in the first term of the
expression for $M$ in~(\ref{K_snova}), we have%
\begin{align}
&  Q_{1}\equiv\frac{1}{q}\int_{0}^{\left\vert \cos\theta\right\vert }J_{R_{0}%
}^{\infty}\xi_{0}d\xi_{0}\nonumber\\
&  =\int_{0}^{\left\vert \cos\theta\right\vert }\left[  \int_{\xi_{0}}%
^{\infty}\left(  1+\beta\eta^{\frac{2}{3}}\right)  \exp\left(  -\ \frac
{3\beta\eta^{\frac{2}{3}}}{2}\right)  \frac{d\eta}{\eta}\right]  \frac{\xi
_{0}d\xi_{0}}{\sqrt{\eta^{2}-\xi_{0}^{2}}}, \label{Q_1}%
\end{align}
where the dimensionless variables $\eta$, $\beta$, and $q$ are defined by
equation~(\ref{beta_snova}). Changing the order of integration, we obtain%
\begin{align*}
&  Q_{1}=\int_{0}^{\left\vert \cos\theta\right\vert }\left(  1+\beta
\eta^{\frac{2}{3}}\right)  \exp\left(  -\ \frac{3\beta\eta^{\frac{2}{3}}}%
{2}\right)  \frac{d\eta}{\eta}\ \int_{0}^{\eta}\frac{\xi_{0}d\xi_{0}}%
{\sqrt{\eta^{2}-\xi_{0}^{2}}}\\
&  +\int_{\left\vert \cos\theta\right\vert }^{\infty}\left(  1+\beta
\eta^{\frac{2}{3}}\right)  \exp\left(  -\ \frac{3\beta\eta^{\frac{2}{3}}}%
{2}\right)  \frac{d\eta}{\eta}\ \int_{0}^{\left\vert \cos\theta\right\vert
}\frac{\xi_{0}d\xi_{0}}{\sqrt{\eta^{2}-\xi_{0}^{2}}}.
\end{align*}
These integrations yield%
\begin{align}
&  Q_{1}=\sqrt{\frac{2\pi}{3\beta^{3}}}\operatorname{erf}\left(  \sqrt
{\frac{3\beta}{2}}\left\vert \cos\theta\right\vert ^{\frac{1}{3}}\right)
+J_{1}\nonumber\\
&  -\left(  \left\vert \cos\theta\right\vert ^{\frac{2}{3}}+\frac{2}{\beta
}\right)  \left\vert \cos\theta\right\vert ^{\frac{1}{3}}\exp\left(
-\ \frac{3\beta\left\vert \cos\theta\right\vert ^{\frac{2}{3}}}{2}\right)  ,
\label{First_term}%
\end{align}
where $\operatorname{erf}(x)=(2/\sqrt{\pi})\int_{0}^{x}e^{-x^{2}}dx$ is the
standard error-function and%
\begin{equation}
J_{1}=\int_{\left\vert \cos\theta\right\vert }^{\infty}\left(  1+\beta
\eta^{\frac{2}{3}}\right)  \exp\left(  -\ \frac{3\beta\eta^{\frac{2}{3}}}%
{2}\right)  \left(  1-\sqrt{1-\frac{\left\vert \cos\theta\right\vert ^{2}%
}{\eta^{2}}}\right)  d\eta. \label{J_1_snova}%
\end{equation}
The integral $J_{1}$ cannot be taken exactly, but directly below we obtain its
approximate value. However, even without doing this, one can easily verify
that equations~(\ref{First_term}) and (\ref{J_1_snova}) provide both the
correct limit of short distances, $\lim_{\beta\rightarrow0}Q_{1}=\pi\left\vert
\cos\theta\right\vert /2$, and the large-distance asymptotics of $\beta\gg1$,
$Q_{1}\approx(2\pi/3)^{1/2}\beta^{-3/2}$.

Now we find an approximate expression for $J_{1}$ by constructing a proper
analytic interpolation between two limiting cases. For small $\beta$, we have%
\begin{equation}
\left.  J_{1}\right\vert _{\beta\rightarrow0}=\left(  \frac{\pi}{2}-1\right)
\left\vert \cos\theta\right\vert . \label{J_1_beta->0}%
\end{equation}
In the opposite limit of large $\beta$, the major contribution to the integral
$J_{1}$ is made in the small vicinity of the lower integration limit. This
yields the following asymptotics,%
\begin{equation}
J_{1}\approx\left(  1-\sqrt{\frac{\pi}{2\beta\left\vert \cos\theta\right\vert
^{\frac{2}{3}}}}\right)  \exp\left(  -\ \frac{3\beta\left\vert \cos
\theta\right\vert ^{\frac{2}{3}}}{2}\right)  \left\vert \cos\theta\right\vert
. \label{J_1_beta>>1}%
\end{equation}
Interpolating between equations ~(\ref{J_1_beta->0}) and (\ref{J_1_beta>>1})
as%
\begin{equation}
J_{1}\approx\left(  1-\frac{\left(  4-\pi\right)  \sqrt{2\pi}}{2\sqrt
{2\pi+\left(  4-\pi\right)  ^{2}\beta\left\vert \cos\theta\right\vert
^{\frac{2}{3}}}}\right)  \exp\left(  -\ \frac{3\beta\left\vert \cos
\theta\right\vert ^{\frac{2}{3}}}{2}\right)  \left\vert \cos\theta\right\vert
, \label{J_1_approxy}%
\end{equation}
we obtain a reasonably good approximation for $J_{1}$, valid in the entire
range of $\beta$. Even in the worst case of $\beta\left\vert \cos
\theta\right\vert ^{\frac{2}{3}}\sim6$, the mismatch between the actual
integral value and this approximation is only $\simeq6\%$.

Combining equations~(\ref{First_term}) with (\ref{J_1_approxy}), for the
double integral $Q_{1}$ defined by equation~(\ref{Q_1}), to a good accuracy we
obtain%
\begin{align}
&  Q_{1}\approx\sqrt{\frac{2\pi}{3\beta^{3}}}\ \operatorname{erf}\left(
\sqrt{\frac{3\beta}{2}}\left\vert \cos\theta\right\vert ^{\frac{1}{3}}\right)
\nonumber\\
&  -\left[  \frac{\left(  4-\pi\right)  \sqrt{2\pi}\left\vert \cos
\theta\right\vert }{2\sqrt{2\pi+\beta\left(  4-\pi\right)  ^{2}\left\vert
\cos\theta\right\vert ^{\frac{2}{3}}}}+\frac{2\left\vert \cos\theta\right\vert
^{\frac{1}{3}}}{\beta}\right]  \exp\left(  -\ \frac{3\beta\left\vert
\cos\theta\right\vert ^{\frac{2}{3}}}{2}\right)  . \label{First_term_final}%
\end{align}

\paragraph{Second term in\ equation~(\ref{K_snova}).}

Now we calculate the integral%
\begin{align}
&  Q_{2}\equiv\frac{1}{q}\int_{0}^{1}J_{R_{0}}^{R}\xi_{0}d\xi_{0}\nonumber\\
&  =\int_{0}^{\left\vert \cos\theta\right\vert }\left[  \int_{\xi_{0}}%
^{1}\left(  1+\beta\eta^{\frac{2}{3}}\right)  \exp\left(  -\ \frac{3\beta
\eta^{\frac{2}{3}}}{2}\right)  \frac{d\eta}{\eta}\right]  \frac{\xi_{0}%
d\xi_{0}}{\sqrt{\eta^{2}-\xi_{0}^{2}}}. \label{Q_2}%
\end{align}
Unlike $Q_{1}$, this integral can be calculated exactly. Indeed, changing the
order of integration, we obtain%
\begin{align}
&  Q_{2}=\int_{0}^{1}\left(  1+\beta\eta^{\frac{2}{3}}\right)  \exp\left(
-\ \frac{3\beta\eta^{\frac{2}{3}}}{2}\right)  \frac{d\eta}{\eta}\ \int%
_{0}^{\eta}\frac{\xi_{0}d\xi_{0}}{\sqrt{\eta^{2}-\xi_{0}^{2}}}\nonumber\\
&  =\int_{0}^{1}\left(  1+\beta\eta^{\frac{2}{3}}\right)  \exp\left(
-\ \frac{3\beta\eta^{\frac{2}{3}}}{2}\right)  d\eta\nonumber\\
&  =\sqrt{\frac{2\pi}{3\beta^{3}}}\ \operatorname{erf}\left(  \sqrt
{\frac{3\beta}{2}}\right)  -\left(  1+\frac{2}{\beta}\right)  \exp\left(
-\ \frac{3\beta}{2}\right)  . \label{Second_term_final}%
\end{align}

\paragraph{Density along the major axis.}

Now we consider two particular positions along the major axis: strictly behind
the meteoroid ($\theta=0$) and strictly in front of it ($\theta=\pi$). In both
these positions, we have $\left\vert \cos0\right\vert =1$, so that the third
and fourth terms in the RHS\ of equation~(\ref{K_snova}) become zero. The
combination of the two first terms there is given by%
\begin{equation}
\pi\left\vert \cos\theta\right\vert \int_{0}^{\left\vert \cos\theta\right\vert
}J_{R_{0}}^{\infty}\xi_{0}d\xi_{0}+\pi\left(  \cos\theta\right)  \int_{0}%
^{1}J_{R_{0}}^{R}\xi_{0}d\xi_{0}=\pi q\left(  \left\vert \cos\theta\right\vert
Q_{1}+\left(  \cos\theta\right)  Q_{2}\right)  \label{combination}%
\end{equation}
As a result, at the major axis behind the meteoroid we obtain
\begin{align}
&  \left.  n^{\left(  2\right)  }\right\vert _{\theta=0}=\frac{8\pi
r_{\mathrm{M}}^{2}n_{0}n_{\mathrm{A}}}{R\sqrt{3}}\left(  1+\frac{m}{m_{\beta}%
}\right)  G_{\mathrm{ion}}\left\{  2\sqrt{\frac{2\pi}{3}}\frac{\lambda
_{T}^{(1)}}{R}\operatorname{erf}\left[  \sqrt{\frac{3}{2}}\left(  \frac
{R}{\lambda_{T}^{(1)}}\right)  ^{\frac{2}{3}}\right]  \right. \nonumber\\
&  \left.  -\left[  \frac{\left(  4-\pi\right)  \sqrt{2\pi}}{2\sqrt
{2\pi+\left(  4-\pi\right)  ^{2}(R/\lambda_{T}^{(1)})^{\frac{2}{3}}}%
}+1+4\left(  \frac{\lambda_{T}^{(1)}}{R}\right)  ^{\frac{2}{3}}\right]
\exp\left[  -\ \frac{3}{2}\left(  \frac{R}{\lambda_{T}^{(1)}}\right)
^{\frac{2}{3}}\right]  \right\}  . \label{n^(2)_theta=0}%
\end{align}
Similarly, at the major axis in front of of the meteoroid we obtain
\begin{figure}[h]
\centering
\includegraphics[width=30pc]{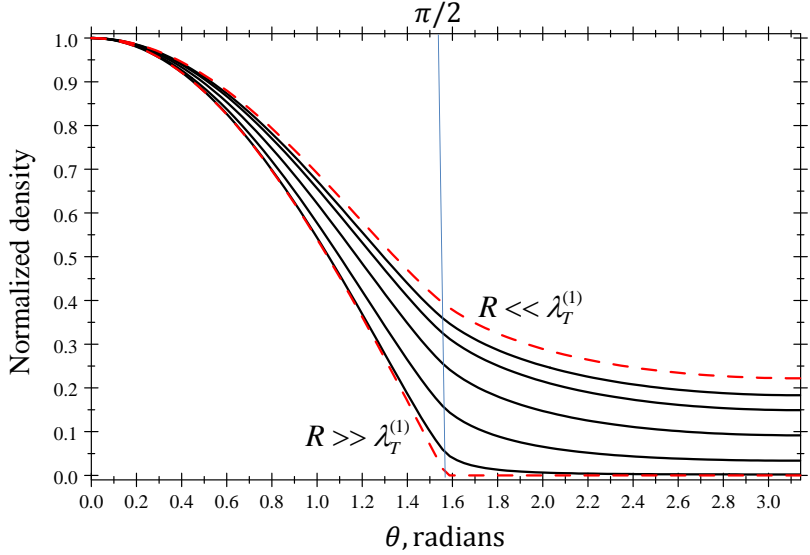}
\caption{Density versus the polar angle $\theta$ for different distances
(black solid curves from the top to the bottom: $R/\lambda_{T}^{(1)}%
=0.1;0.3;1;3;10$). Red dashed curves show asymptotic solutions given by
equation~(\ref{K_beta>>1}) (the top curve) and by equation~(\ref{n_R<<lambda}%
). All density distributions are normalized to the maximum values strictly
behind the meteoroid, $\theta=0$ (see
Fig.~\ref{Fig:Density_vs_Radius_logarithm}).}
\label{Fig:M_versus_theta}%
\end{figure}
\begin{figure}[h]
\centering
\includegraphics[width=30pc]{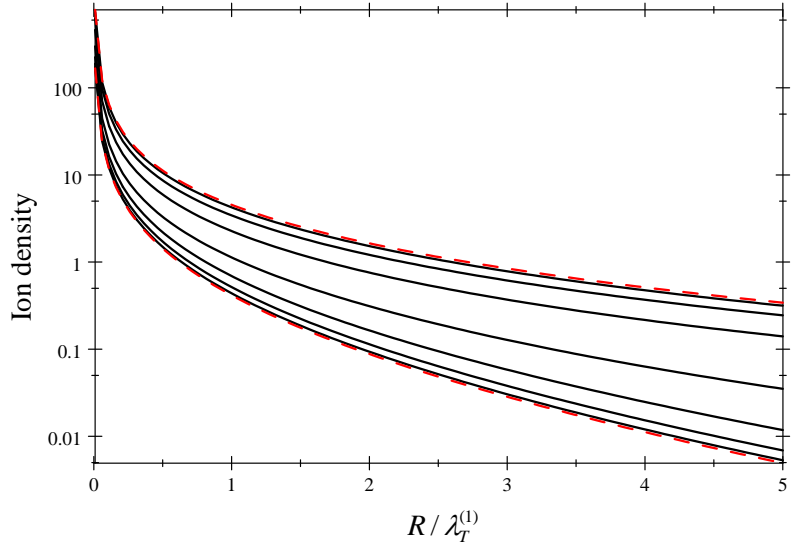}
\caption{Normalized density versus the relative radial distance,
$R/\lambda_{T}^{(1)}$, for several equally separated angles $\theta$. The
normalization corresponds to the factor $M$ with $q\rightarrow\lambda
_{T}^{(1)}/R$. Red dashed curves for $\theta=0$ (the top curve) and for
$\theta=\pi$ (the bottom curve) are described analytically by
equations~(\ref{n^(2)_theta=0}) and (\ref{n^(2)_theta=pi}), respectively.
Black solid curves show $M(R,\theta)$ obtained by combining analytic
equations~(\ref{Q_1}) and (\ref{Q_2}) for the two first terms of
equation~(\ref{K_snova}) with numerically calculated two remaining integral
terms. The black solid curves, from the top to the bottom, correspond to
$\theta=\pi/8$; $\pi/4$; $3\pi/8$; $\pi/2$; $5\pi/8$; $3\pi/4;$ and $7\pi/8$,
respectively. The black solid curves with $\theta=\pi/8$ and $\theta=7\pi/8$
are almost indistinguishable from the top and bottom red dashed curves,
$\theta=0,\pi$.}%
\label{Fig:Density_vs_Radius_logarithm}
\end{figure}
\begin{align}
&  \left.  n^{\left(  2\right)  }\right\vert _{\theta=\pi}=\frac{8\pi
r_{\mathrm{M}}^{2}n_{0}n_{\mathrm{A}}}{R\sqrt{3}}\left(  1+\frac{m}{m_{\beta}%
}\right)  G_{\mathrm{ion}}\nonumber\\
&  \times\left[  1-\frac{\left(  4-\pi\right)  \sqrt{2\pi}}{2\sqrt
{2\pi+\left(  4-\pi\right)  ^{2}(R/\lambda_{T}^{(1)})^{\frac{2}{3}}}}\right]
\exp\left[  -\ \frac{3}{2}\left(  \frac{R}{\lambda_{T}^{(1)}}\right)
^{\frac{2}{3}}\right]  . \label{n^(2)_theta=pi}%
\end{align}

Figure~\ref{Fig:Density_vs_Radius_logarithm} shows these two radial
dependencies with $\cos\theta=1$ ($\theta=0$) and $\cos\theta=-1$ ($\theta
=\pi$) are actually representative for slightly more general $\cos\theta$. The
radial-distance dependencies for positive and negative $\cos\theta$ look
qualitatively different, as we discuss below.
\begin{figure}[h]
\centering
\includegraphics[width=30pc]
{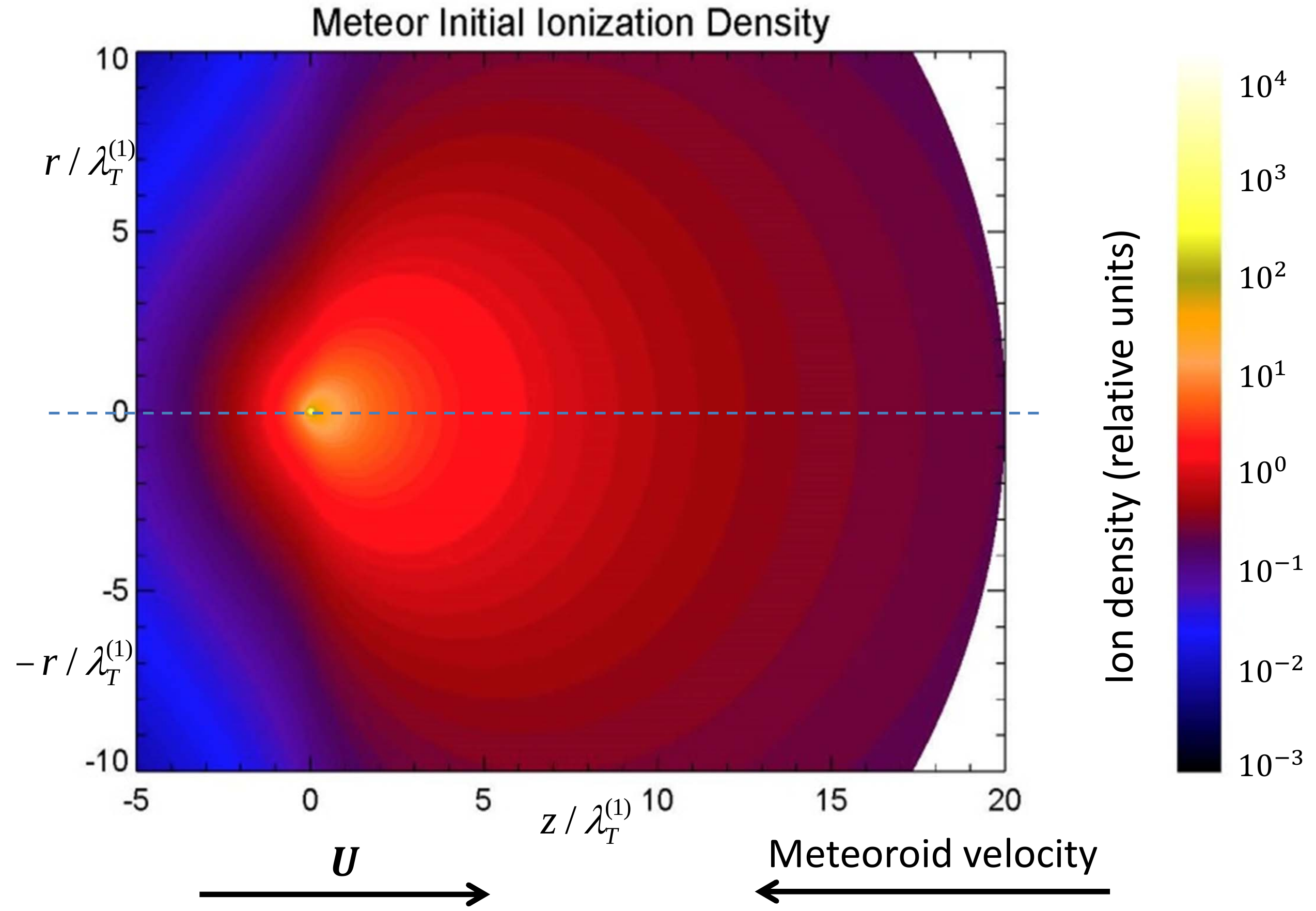}
\caption{Spatial distribution of plasma density around a small meteoroid. The
figure shows a cross-section of an axially symmetric distribution by a plane
that includes the symmetry axis (blue dashed line). The meteoroid is shown by
a small white circle at the symmetry axis. The color coding signifies the
plasma density in relative units; $r$ is the distance to the symmetry axis.}
\label{Fig:3Ddensity}
\end{figure}

Behind the meteoroid, $\cos\theta>0$, the radial dependence of the density
gradually changes from $n^{\left(  2\right)  }\varpropto1/R$ for $R\ll
\lambda_{T}^{(1)}$, as described by equation~(\ref{n_R<<lambda}), to
$n^{\left(  2\right)  }\varpropto1/R^{2}$ for $R\gg\lambda_{T}^{(1)}$, as
described by equation~(\ref{n_R>>lambda}). This change in the power-law radial
dependence of $n^{\left(  2\right)  }$ occurs for the following reason. The
source for the secondary ions are the primary neutral particles, whose density
falls off near the meteoroid roughly as $1/R^{2}$. At a given location $R$,
the number of ions moving in a certain direction is determined by the total
collisional ionization over the preceding segment of the straight-line
trajectory aligned with that direction. For $R\ll\lambda_{T}^{(1)}$, the total
integration over the entire ionization path acquires an additional factor
$\varpropto R$ which gradually transforms $1/R^{2}$ to $1/R$. On the other
hand, for $R\gg\lambda_{T}^{(1)}$ only localized ionization within $R^{\prime
}\lesssim\lambda_{T}^{(1)}$ plays a role, resulting in `saturation' of the
previous additional factor $R$ at a constant value $\sim\lambda_{T}^{(1)}$.
This leaves the $\varpropto1/R^{2}$ dependence of $n^{\left(  2\right)  }$
essentially untouched. This transition works only for locations behind the
meteoroid, $\cos\theta>0$, because almost all freshly born ions have
velocities $\vec{V}$ with positive $\mu=\cos\vartheta$, where $\vartheta$ is
the angle between $\vec{V}$ and $-\vec{U}$, as shown in
Figure~\ref{Fig:Cartoon_reproduced}. Regardless of how far away from the
meteoroid this $R,\theta$-point is located, all preceding straight-line
trajectory segments with $\mu>0$ always cross the near-meteoroid volume
$R^{\prime}\lesssim\lambda_{T}^{(1)}$,

A significantly different situation takes place in front of the meteoroid,
$\cos\theta<0$. At $R\ll\lambda_{T}^{(1)}$, straight-line trajectory segments
with $\mu>0$ also cross a part of the near-meteoroid volume $R^{\prime
}\lesssim\lambda_{T}^{(1)}$. That is why here $\left.  n^{\left(  2\right)
}\right\vert _{\theta=\pi}$ is quite noticeable, although a few times smaller
than $\left.  n^{\left(  2\right)  }\right\vert _{\theta=0}$. On the other
hand, at larger distances, $R\gtrsim\lambda_{T}^{(1)}$, there are almost no
preceding trajectory segments with positive $\mu$ that would cross the near
region of $R^{\prime}\lesssim\lambda_{T}^{(1)}$. These trajectories cross the
regions where the number of the primary particles is itself exponentially
small, so that $\left.  n^{\left(  2\right)  }\right\vert _{\theta=\pi
}\varpropto\exp[-\ (3/2)(R/\lambda_{T}^{(1)})^{2/3}]$. This exponentially
decreasing density remains much less than that behind the meteoroid where
$n^{\left(  2\right)  }$ decreases largely by a power law.

\paragraph{General case.\label{General case_again}}

For the general case of $R\sim\lambda_{T}^{(1)}$ with $\cos\theta\neq\pm1$, we
were unable to find acceptable analytic approximations for the two last
(integral) terms in the RHS\ of equation~(\ref{K_snova}). Therefore, we will
integrate those 1D integrals numerically.

Now we summarize the entire expression for $n^{\left(  2\right)  }$ by
combining equation~(\ref{n^(2)_prome_2}) with (\ref{K_snova}), where in the
first two terms the integrals $\int_{0}^{\left\vert \cos\theta\right\vert
}J_{R_{0}}^{\infty}\xi_{0}d\xi_{0}=qQ_{1}$ and $\int_{0}^{1}J_{R_{0}}^{R}%
\xi_{0}d\xi_{0}=qQ_{2}$ were calculated with $Q_{1,2}$ explicitly given by
equations~(\ref{First_term_final}) and (\ref{Second_term_final}) (as it was
done above when calculating the ion density at the major axis). For the two
remaining integral terms in equation~(\ref{K_snova}) we use the approximation
for $J_{R_{0}}^{\infty}$ given by equation~(\ref{J_R_)_approx_dimensionless}).
This gives%
\begin{align}
n^{\left(  2\right)  }  &  =\frac{8\pi r_{\mathrm{M}}^{2}n_{0}n_{\mathrm{A}}%
}{\sqrt{3}R}\left(  1+\frac{m}{m_{\beta}}\right)  G_{\mathrm{ion}%
}(U)\nonumber\\
&  \times\left[  f_{1}\left(  R,\cos\theta\right)  \left\vert \cos
\theta\right\vert +f_{2}\left(  R\right)  \cos\theta+f_{3}\left(  R,\cos
\theta\right)  \right]  , \label{n^(2)_general}%
\end{align}
where%
\begin{align}
&  f_{1}\left(  R,\cos\theta\right)  =\frac{\lambda_{T}^{(1)}}{R}\sqrt
{\frac{2\pi}{3}}\operatorname{erf}\left[  \sqrt{\frac{3}{2}}\left(  \frac
{R}{\lambda_{T}^{(1)}}\right)  ^{\frac{1}{3}}\left\vert \cos\theta\right\vert
^{\frac{1}{3}}\right] \nonumber\\
&  -\left[  \frac{\left(  4-\pi\right)  \left\vert \cos\theta\right\vert
}{2\sqrt{1+\left.  \left(  4-\pi\right)  ^{2}(R/\lambda_{T}^{(1)})^{\frac
{2}{3}}\left\vert \cos\theta\right\vert ^{\frac{2}{3}}\right/  (2\pi)}%
}+2\left(  \frac{\lambda_{T}^{(1)}}{R}\right)  ^{\frac{2}{3}}\left\vert
\cos\theta\right\vert ^{\frac{1}{3}}\right] \nonumber\\
&  \times\exp\left[  -\ \frac{3\left\vert \cos\theta\right\vert ^{\frac{2}{3}%
}}{2}\left(  \frac{R}{\lambda_{T}^{(1)}}\right)  ^{\frac{2}{3}}\right]  ,
\label{f_1}%
\end{align}

\begin{align}
f_{2}\left(  R\right)   &  =\frac{\lambda_{T}^{(1)}}{R}\sqrt{\frac{2\pi}{3}%
}\operatorname{erf}\left[  \sqrt{\frac{3}{2}}\left(  \frac{R}{\lambda
_{T}^{(1)}}\right)  ^{\frac{1}{3}}\right] \nonumber\\
&  -\left[  1+2\left(  \frac{\lambda_{T}^{(1)}}{R}\right)  ^{\frac{2}{3}%
}\right]  \exp\left[  -\ \frac{3}{2}\left(  \frac{R}{\lambda_{T}^{(1)}%
}\right)  ^{\frac{2}{3}}\right]  , \label{f_2}%
\end{align}

\begin{align}
f_{3}(R,\cos\theta)  &  =\int_{\left\vert \cos\theta\right\vert }^{1}%
\sqrt{1+\frac{2\xi_{0}^{\frac{2}{3}}}{\pi}\left(  \frac{R}{\lambda_{T}^{(1)}%
}\right)  ^{\frac{2}{3}}}\exp\left[  -\ \frac{3\xi_{0}^{\frac{2}{3}}}%
{2}\left(  \frac{R}{\lambda_{T}^{(1)}}\right)  ^{\frac{2}{3}}\right]
\nonumber\\
&  \times\sqrt{\frac{\xi_{0}^{2}-\cos^{2}\theta}{1-\xi_{0}^{2}}}\ d\xi
_{0}\nonumber\\
&  +\left\vert \cos\theta\right\vert \int_{\left\vert \cos\theta\right\vert
}^{1}\sqrt{1+\frac{2\xi_{0}^{\frac{2}{3}}}{\pi}\left(  \frac{R}{\lambda
_{T}^{(1)}}\right)  ^{\frac{2}{3}}}\exp\left[  -\ \frac{3\xi_{0}^{\frac{2}{3}%
}}{2}\left(  \frac{R}{\lambda_{T}^{(1)}}\right)  ^{\frac{2}{3}}\right]
\nonumber\\
&  \times\arcsin\frac{\sqrt{1-\xi_{0}^{2}}\left\vert \cos\theta\right\vert
}{\xi_{0}\sqrt{1-\cos^{2}\theta}}\ d\xi_{0}. \label{f_3}%
\end{align}
For the isotropic differential cross-section the mean free path defined
equation by equation~(\ref{lambda_T^(1)}) reduces to
\begin{equation}
\lambda_{T}^{(1)}=\frac{1}{4\pi Un_{\mathrm{A}}G(U)}\left(  \frac
{T_{\mathrm{M}}}{m_{\mathrm{M}}}\right)  ^{1/2}. \label{lambda_new}%
\end{equation}

Figures~\ref{Fig:M_versus_theta}, \ref{Fig:Density_vs_Radius_logarithm}, and
\ref{Fig:3Ddensity}\ illustrate the general $\theta,R$-dependences of
$n^{\left(  2\right)  }$.

As might be expected, in Figure~\ref{Fig:M_versus_theta} the normalized curves
with intermediate values of $R/\lambda_{T}^{(1)}$ smoothly and uniformly fill
the gap between the two asymptotic solutions corresponding to the long,
$R\gg\lambda_{T}^{(1)}$, and short, $R\ll\lambda_{T}^{(1)}$, distances, as
described by equations~(\ref{n_R>>lambda}) and (\ref{n_R<<lambda}). Similarly,
in Figure~\ref{Fig:Density_vs_Radius_logarithm} the curves with intermediate
values of $\cos\theta$ fill the gap between the solutions strictly in front of
the meteoroid ($\cos\theta=-1$) and behind it ($\cos\theta=1$), as described
by equations~(\ref{n^(2)_theta=pi}) and (\ref{n^(2)_theta=0}).

Figure~\ref{Fig:3Ddensity} shows the entire 3D structure of the ion density in
color coding. Since the spatial distribution of the plasma density is axially
symmetric, this figure shows a meridional cross-section that includes the
major axis. Behind the meteoroid, at $z\equiv R\cos\theta\gg\lambda_{T}^{(1)}$
the presented distribution may be inaccurate because it does not properly
describe the extended trail formed lagging behind multiply scattered ions.
However, around the meteoroid this distribution should be reasonably accurate.
Figure~\ref{Fig:3Ddensity} shows a significant span in the density values
(seven orders of magnitude). Notice visible bulges in constant density
contours in front of the meteoroid. This feature has not been predicted in
previous PIC simulations \citep{Dyrud:Plasma2008EMaP,Dyrud:Plasma2008AdSpR}.

\section{Discussion\label{Discussion}}

The spatial structure of the near-meteoroid plasma shown in
Figure~\ref{Fig:3Ddensity} scales with the collisional mean free path of the
primary (ablated) particles, $\lambda_{T}^{(1)}$. The latter, in turn, scales
with the altitude($h$)-dependent atmospheric density as $\lambda_{T}%
^{(1)}\propto1/n_{\mathrm{A}}(h)$. The frontal edge of the meteoroid plasma
has a bulge located around the major axis at the distance $\sim\lambda
_{T}^{(1)}$ from the meteoroid.

The most striking feature of the meteor-plasma spatial structure is not the
angular dependence of $n^{(2)}$ but rather the fact that near the meteoroid,
$R\lesssim\lambda_{T}^{(1)}$, the plasma density behaves as $n^{\left(
2\right)  }\simeq F(\theta)/R$. The singular $1/R$ dependence holds almost to
the meteoroid surface, $R\sim r_{\mathrm{M}}$. This fact has serious
implications for head-echo observations and for the corresponding modeling of
the radar wave propagation.

To analyze radio wave propagation through the dense meteoroid plasma with
$n_{e}\approx n_{i}\approx n^{(2)}\gg n_{\mathrm{A}}$, we apply the simplest
criterion of geometrical optics applicable (the WKB approximation):
\begin{equation}
\frac{\lambda_{0}}{2\pi}\ \frac{\left\vert \nabla\tilde{n}\right\vert
}{|\tilde{n}|^{2}}\ll1 \label{WKB}%
\end{equation}
\citep{Ginzburg:Propagation71}. Here $\lambda_{0}=c/f$ is the vacuum wavelength
with the frequency $f$, and index of refraction (neglecting the geomagnetic
field) given by $\tilde{n}=\left(  1-n_{e}/n_{\mathrm{cr}}\right)  ^{1/2}$,
where%
\begin{equation}
n_{\mathrm{cr}}=\frac{4\pi^{2}\epsilon_{0}m_{e}f^{2}}{e^{2}}\approx
1.24\times10^{4}%
\operatorname{cm}%
^{-3}\left(  \frac{f}{1%
\operatorname{MHz}%
}\right)  ^{2} \label{n_cr}%
\end{equation}
is the critical plasma density corresponding to $\tilde{n}=0$.

Most of small meteoroids have a radius within the hundreds-of-microns range,
i.e., many orders of magnitude less than the mean free path, $\lambda
_{T}^{(1)}$. This means that the radar beam with the wave frequency in the
tens-of-MHz range (e.g., Jicamarca with $f=50~$MHz) will almost certainly
cross the boundary $R=R_{\mathrm{cr}}(\theta)$ where $n_{\mathrm{cr}}%
=n_{e}\approx n^{\left(  2\right)  }(R_{\mathrm{cr}},\theta)$. Under the
geometric-optics approximation, the boundary $R\approx R_{\mathrm{cr}}%
(\theta)$ represents the wave reflection level.

According to equation~(\ref{WKB}), geometric optics applies beyond the wave
reflection level of $\tilde{n}=0$ and, for $n_{e}\propto1/R$, where
\begin{equation}
R\gg\sqrt{\frac{\lambda_{0}R_{\mathrm{cr}}}{4\pi|\tilde{n}|^{3}}}.
\label{proximity}%
\end{equation}
For small radii, $R\ll R_{\mathrm{cr}}$, for which $|\tilde{n}|\approx
(R_{\mathrm{cr}}/R)^{1/2}\gg1$, this requires $R\gg\lambda_{0}^{2}/(16\pi
^{2}R_{\mathrm{cr}})$, meaning that geometric optics applicability will
necessarily fail in the close proximity to the meteoroid.

If $R_{\mathrm{cr}}\gg\lambda_{0}/(4\pi)$ then the radio wave propagating from
larger to smaller $R$ will reach the reflection level of the plasma density,
$R=R_{\mathrm{cr}}$, well before reaching the volume where the WKB becomes
invalid. This corresponds to normal overdense reflection of the radar wave, so
that the $1/R$ dependence of $n^{\left(  2\right)  }$ will not prevent using
the regular geometric-optics technique. If, however, $R_{\mathrm{cr}}%
\lesssim\lambda_{0}$ then the applicability of the geometric-optics
propagation breaks down before the radio wave reaches its reflection level.
This requires modeling the radar wave propagation using full Maxwell's
equations \citep{Marshall:FDTD15}. The characteristics of the radar signal
obtained in this way may differ from those obtained in the framework of the
regular ray tracing.

To conclude this discussion, we note that in the underlying kinetic theory we
have neglected the effect of fields on the ion collisionless motion. Directly
near the meteoroid, where the calculated density sharply increases with
decreasing $R$ this assumption may become invalid. This is especially true
within the Debye length from the meteoroid surface, where one may expect a
significant net positive charge formed by slow ions after fast electrons moved
away from that region. This charge separation can create a significant
potential barrier for electrons that, at the same time, will accelerate ions
away from the meteoroid. This may modify the $1/R$ rate near the meteoroid
surface. However, the relevant distances are typically much smaller than the
radar wavelength, so that this localized modification should not affect the
overall radio wave propagation and formation of the radar head echo.

\section{Conclusions\label{Conclusions}}

Based on the kinetic theory developed in the companion paper
\citep{Dimant:Formation1_17}, we have calculated the spatial distribution of
the meteor plasma responsible for the radar head echo.

The underlying theory assumes that (1) plasma electrons obey the Boltzmann
distribution; (2) most plasma ions originate from the ablated meteoroid
material after exactly one ionizing collision, while further collisions can be
neglected; (3) the ion motion between collisions is largely unaffected by
fields. In order to make plasma density calculations easier, in this paper we
have additionally assumed the isotropic differential cross-section of the
meteor-particle ionization.

Equations~(\ref{n^(2)_general}) to (\ref{f_3}) describe the entire spatial
distribution of the near-meteoroid plasma density.
Figures~\ref{Fig:M_versus_theta}, \ref{Fig:Density_vs_Radius_logarithm}, and
\ref{Fig:3Ddensity}\ illustrate the solution. The major feature of the
near-meteoroid plasma density spatial distribution is its dominant $1/R$
dependence in most of the plasma. This quasi-singular behavior makes the radar
penetrated near-meteoroid plasma to be overdense at locations sufficiently
close to the meteoroid. This and other features are important for modeling the
radar head echo. In near future, we are planning to employ the obtained plasma
density distribution for such modeling.

\section*{Acknowledgments}

This work was supported by NSF Grant AGS-1244842.


\end{document}